\documentclass[
]{ceurart}

\usepackage{listings}
\usepackage{graphicx}
\usepackage{subcaption}
\usepackage{hyperref}
\begin{document}

\copyrightyear{2023}
\copyrightclause{Copyright for this paper by its authors.
  Use permitted under Creative Commons License Attribution 4.0
  International (CC BY 4.0).}

\conference{Proceedings of 4th International Workshop on Human-Centred Learning Analytics (HCLA) co-located with the 13th International Learning
Analytics and Knowledge Conference (LAK2023), Virtual, March 13, 2023.}
\title{Towards a Flexible User Interface for `Quick and Dirty' Learning Analytics Indicator Design}




\author{Shoeb Joarder}[%
email=shoeb.joarder@uni-due.de,
]

\author{Mohamed Amine Chatti}[%
email=mohamed.chatti@uni-due.de,
]

\author{Seyedemarzie Mirhashemi}[%
email=seyedemarzie.mirhashemi@stud.uni-due.de,
]

\author{Qurat Ul Ain}[%
email=qurat.ain@stud.uni-due.de,
]
\address{Social Computing Group, Faculty of Engineering, University of Duisburg-Essen, 47048 Duisburg, Germany}



\begin{abstract}
Research on Human-Centered Learning Analytics (HCLA) has provided demonstrations of a successful co-design process for LA tools with different stakeholders. However, there is a need for `quick and dirty' methods to allow the low-cost design of LA indicators. Recently, Indicator Specification Cards (ISC) have been proposed to help different learning analytics stakeholders co-design indicators in a systematic manner. In this paper, we aim at improving the user experience, flexibility, and reliability of the ISC-based indicator design process. To this end, we present the development details of an intuitive and theoretically-sound ISC user interface that allows the low-cost design of LA indicators. Further, we propose two approaches to support the flexible design of indicators, namely a task-driven approach and a data-driven approach.
\end{abstract}

  
\maketitle

\section{Introduction and Background}
In learning analytics (LA) systems, the learners' interaction data collected from various learning environments is typically shown in dashboards through indicator visualizations. In order to support learning and teaching, various LA dashboards and indicators were suggested \cite{bodily2017review,schwendimann2016perceiving}. However, these are often designed without reference to Human-Computer Interaction (HCI) and information visualization guidelines and theories \cite{gavsevic2015let, jivet2018license, klerkx2017learning}. Furthermore, these indicators are typically not well suited to users' needs and expectations \cite{chatti2019perla, gavsevic2016we}. Recently, Human-Centered Learning Analytics (HCLA) has proven to be important in the LA community and was considered as a key to user acceptance and adoption of LA systems. HCLA is an approach that emphasizes the human factors of LA and brings HCI to LA, to involve stakeholders throughout the LA process \cite{buckingham2019human,chatti2021designing, dimitriadis2021human, sarmiento2022participatory}. In general, research on HCLA encourages active user involvement in the LA design process and provides demonstrations of successful co-design process for LA tools with different stakeholders \cite{dollinger2019working,holstein2019co,alvarez2020deck}. However, the reported case studies were focused on the participatory design of LA tools and platforms (macro design level) rather than the systematic design of the underlying indicators (micro design level). To fill this gap, \citet{chatti2020design} proposed the Human-Centered Indicator Design (HCID) and Indicator Specification Cards (ISC) to support the systematic co-design of LA indicators.
\subsection{Human-Centered Indicator Design}
Human-Centered Indicator Design (HCID) is an HCLA approach that involves users in the systematic design of LA indicators that fit their needs \cite{chatti2020design}. HCID brings together Norman's Human-Centered Design (HCD) process \cite{norman2013design} and Munzner's what-why-how visualization framework \cite{munzner2014visualization} to provide a theory-informed method to design effective LA indicators systematically. The HCID requires the users to take four iterative stages that help develop the right indicator: (1) Define Goal/Question, (2) Ideate, (3) Prototype, and (4) Test. In order to actively involve users in the \textit{Ideate} phase of the HCID process, the authors proposed to use Indicator Specification Cards (ISC).
\subsection{Indicator Specification Cards}
\begin{figure}
    \centering
    \begin{minipage} {0.48\textwidth}
        \centering
        \includegraphics[width=0.9\textwidth]{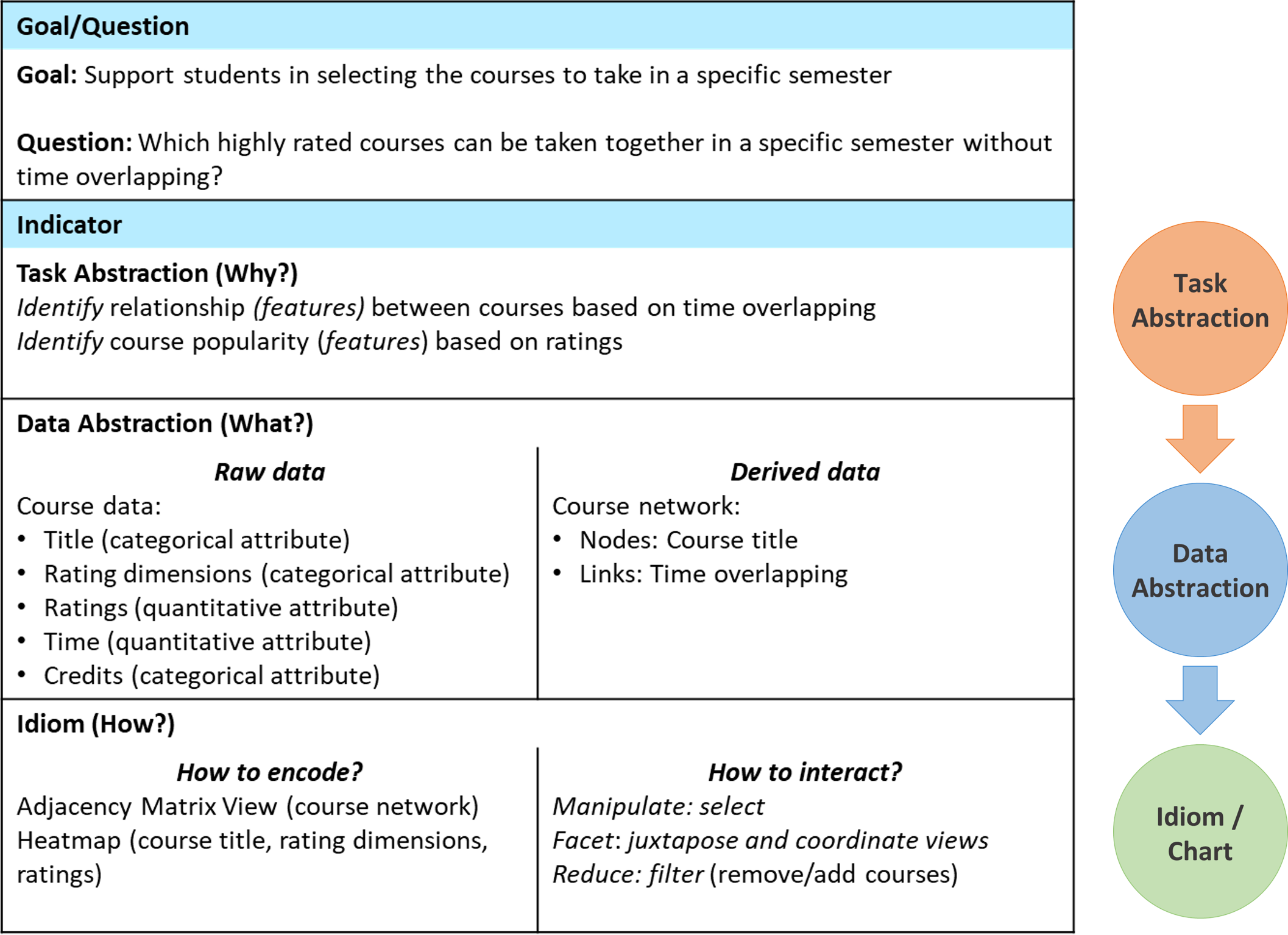}
        \caption{ISC example \cite{chatti2020design}}
        \label{fig:isc-example}
    \end{minipage} \hfill
    \begin{minipage} {0.48\textwidth}
        \centering
        \begin{minipage} {0.87\textwidth}
            \centering
            \includegraphics[width=1\textwidth]{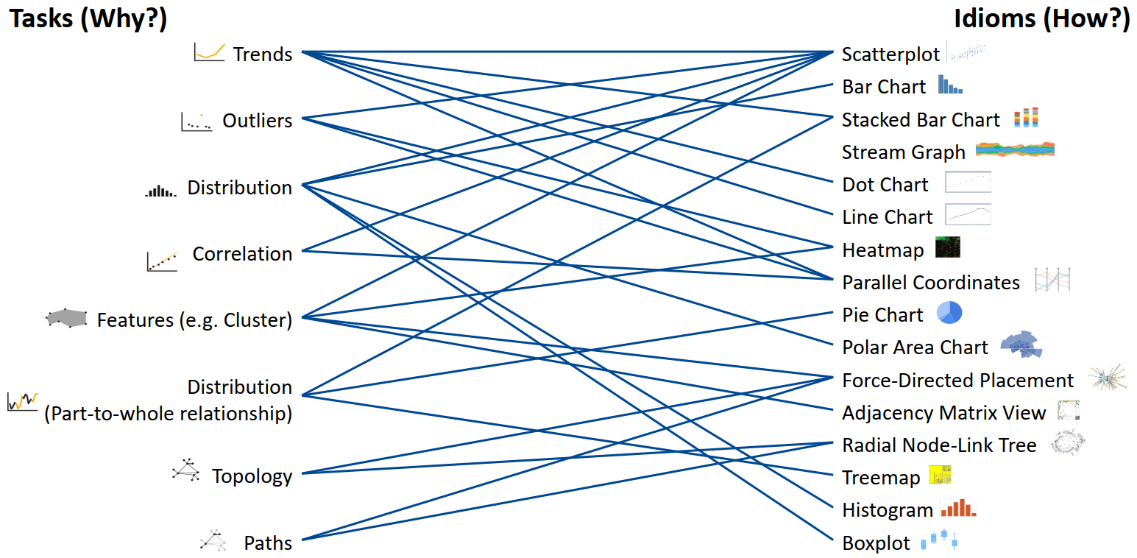}
            \caption{InfoVis design guidelines mapping \\ Why? $\to$ How? \cite{chatti2021designing}}
            \label{subfig:infovis-mapping-a}
        \end{minipage}
        
        \begin{minipage} {0.87\textwidth}
            \centering
            \includegraphics[width=1\textwidth]{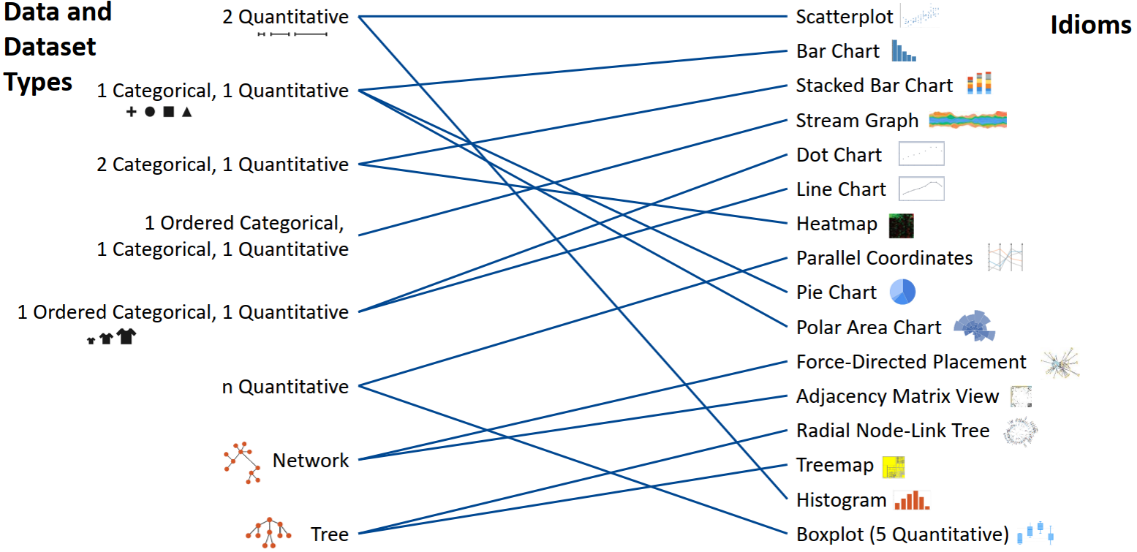}
            \caption{InfoVis design guidelines mapping \\ What? $\to$ How? \cite{chatti2021designing}}
            \label{subfig:infovis-mapping-b}
        \end{minipage}
    \end{minipage}
\end{figure}

Indicator Specification Cards (ISC) represent a theory-informed method that helps different LA stakeholders co-design indicators in a systematic manner \cite{chatti2020design}. An ISC provide a `quick and dirty' method to allow low cost design of LA indicators. It follows the Goal-Question-Indicator (GQI) approach \cite{muslim2017goal} to design LA indicators that meet users' goals and applies information visualization guidelines from Munzner's what-why-how visualization framework \cite{munzner2014visualization}. Concretely, it describes a systematic workflow to get from the \textit{why?} (i.e., user goal/question) to the \textit{how?} (i.e., visualization). It consists of two main parts, namely \textit{Goal/Question} and \textit{Indicator}, as shown in Figure~\ref{fig:isc-example}. The \textit{Goal/Question} part refers to the outcomes of the \textit{Define Goal/Question} stage of the HCID approach. The \textit{Indicator} part is further broken down into three sub-parts, namely \textit{Task Abstraction (Why?)}, \textit{Data Abstraction (What?)}, and \textit{Idiom (How?)}, which reflect the three dimensions of Munzner's what-why-how visualization framework. The information visualization (InfoVis) literature suggests that the \textit{Idioms (How?)} depend heavily on the underlying \textit{Tasks (Why?)} and \textit{Data (What?)} of the visualization and provides guidelines to ``what kind of idioms is more effective for what kind of tasks (mapping Why? $\to$ How?)'' (Figure~\ref{subfig:infovis-mapping-a}) and ``what kind of idioms is more effective for what kind of data (mapping What? $\to$ How?)'' (Figure~\ref{subfig:infovis-mapping-b}). ISCs were used in \cite{chatti2021designing} to co-design LA indicators to support self-regulated learning (SRL) activities of bachelor students attending an introductory Python programming course. The authors experienced that using ISCs was a complex, time-consuming, and resource-intensive task. This is mainly due to the fact that using ISCs imposes one particular sequence of steps to be followed to design the indicators (i.e., \textit{Task Abstraction} $\to$ \textit{Data Abstraction} $\to$ \textit{Idiom/Chart}). Moreover, the authors reported that the step related to the selection of the appropriate idioms/charts turned out to be too complex and at times confusing to the participants, since in general, bachelor students do not have a strong background in information visualization design practices. The participants found the provided mappings (Figure~\ref{subfig:infovis-mapping-a} and Figure~\ref{subfig:infovis-mapping-b}) helpful, but still complex to help choose the right idioms/charts for the task and data at hand.
\section{ISC Design Goals}
To address the limitations in the design of ISCs outlined above, we conducted a qualitative user study with 16 participants (12 learners and four teachers) to develop a more profound knowledge of the thought process of how they would design LA indicators and their expectations from using ISCs to systematically design indicators. Based on the interviews and observations, we gained a better understanding of the end-users' needs, and determined the following design goals for ISCs:

\textbf{(DG1) Intuitive User Interface:} Lack of a User Interface (UI) was a critical issue for users to realize how they would design their indicators using ISCs. The users required an intuitive UI for the ISC that they can interact with to design indicators in a straightforward manner. 

\textbf{(DG2) Flexible sequence of steps:} Users found that following the same predefined sequence of steps (i.e., \textit{Task Abstraction} $\to$ \textit{Data Abstraction} $\to$ \textit{Idiom/Chart}) to be restrictive, and preferred to have a flexible system that would support them in self-defining the order of the steps to be taken to design their indicators.  

\textbf{(DG3) Recommendations of idioms/charts:} The system should be able to provide recommendations of suitable idioms/charts based on chosen data and/or task at hand, following guidelines from the information visualization literature.
\section{ISC Design and Implementation}
We followed the Human-Centered Design (HCD) approach and usability design principles \cite{norman2013design} to systematically design an intuitive ISC UI \textbf{(DG1)}. The ISC UI was designed in three iterations by using low-fidelity prototypes, software prototypes, and high-fidelity prototypes. After each iteration, feedback was addressed in the subsequent design, which was then again evaluated. We designed and implemented two flexible approaches to design indicators \textbf{(DG2)}, namely (1) \textit{task-driven approach} and (2) \textit{data-driven approach}. For both approaches, we integrated recommendations of suitable idioms/charts based on chosen data and/or task at hand \textbf{(DG3)}.
\begin{itemize}
    \item \textbf{Task-driven approach:} Users have a specific goal in mind and based on their goal, they can choose an appropriate task, then choose a recommended idiom/chart, and finally select the appropriate data (i.e., \textit{Task Abstraction} $\to$ \textit{Idiom/Chart} $\to$ \textit{Data Abstraction}).
    \item \textbf{Data-driven approach:} Users have data and based on the data types, users can either choose a recommended idiom/chart directly (i.e., \textit{Data Abstraction} $\to$ \textit{Idiom/Chart}) or choose a task and then choose a recommended idiom/chart (i.e., \textit{Data Abstraction} $\to$ \textit{Task Abstraction} $\to$ \textit{Idiom/Chart}).
\end{itemize}

We created the ISC UI using modern web technologies such as React.JS, Material Design, React-Redux for state management, and ApexCharts.JS for visualizations. The implementation details of the two different approaches are described in the following sections.

\subsection{Task-driven Approach}
\begin{figure}[htpb]
    \begin{center}
        \begin{subfigure}[normla]{0.48\textwidth}
            \includegraphics[width=1\textwidth]{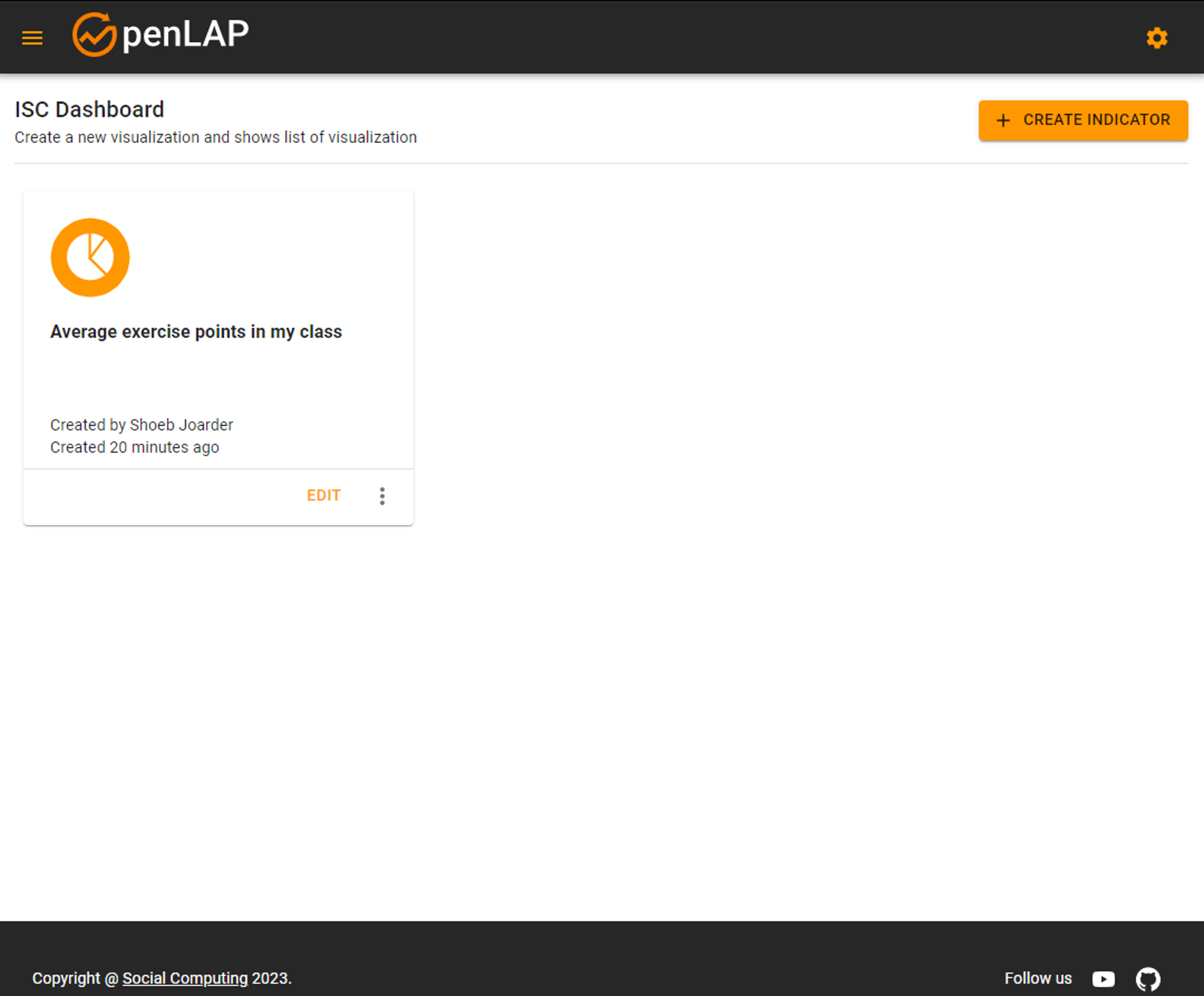}
            \caption{ISC dashboard}
            \label{subfig:a1}
        \end{subfigure}
        ~
        \begin{subfigure}[normla]{0.48\textwidth}
            \includegraphics[width=1\textwidth]{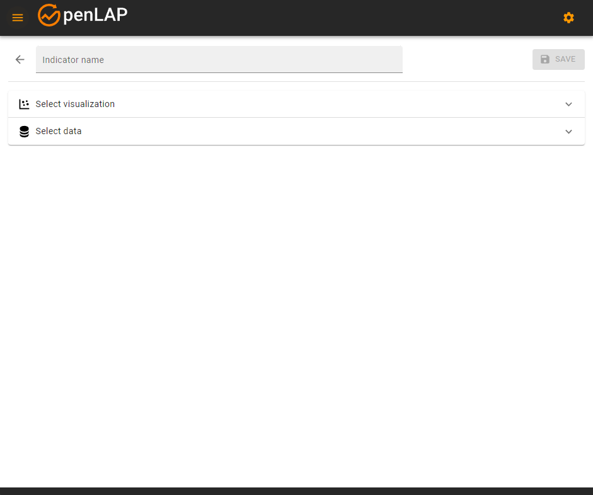}
            \caption{Create Indicator}
            \label{subfig:a2}
        \end{subfigure}
        \\
        \begin{subfigure}[normla]{0.48\textwidth}
            \includegraphics[width=1\textwidth]{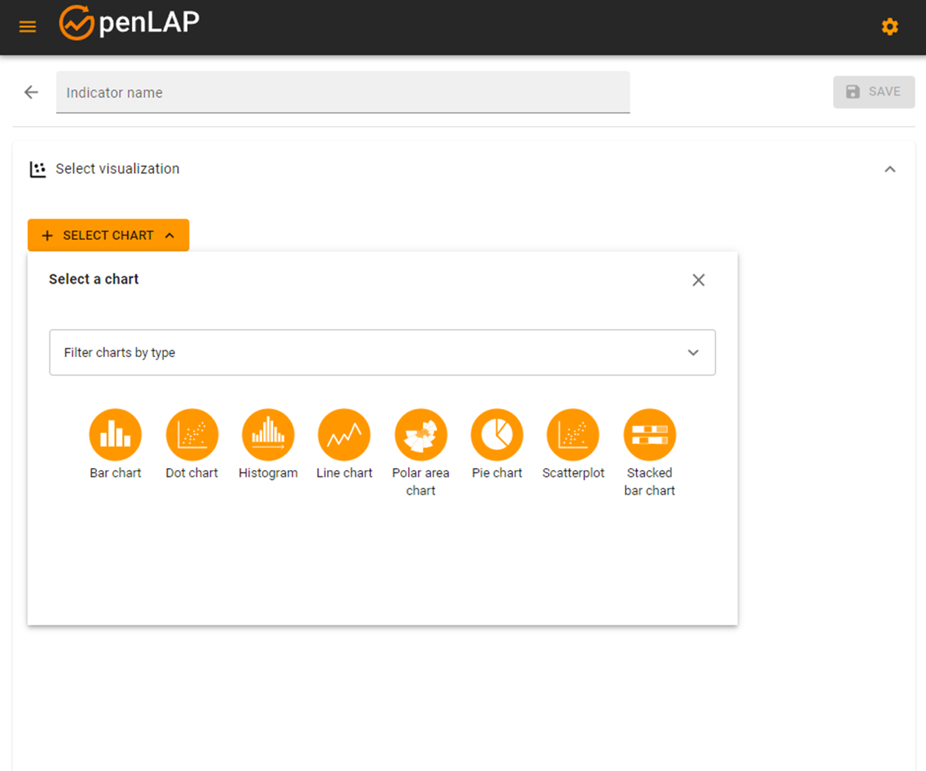}
            \caption{View list of \textit{Idioms}}
            \label{subfig:a3}
        \end{subfigure}
        ~
        \begin{subfigure}[normla]{0.48\textwidth}
            \includegraphics[width=1\textwidth]{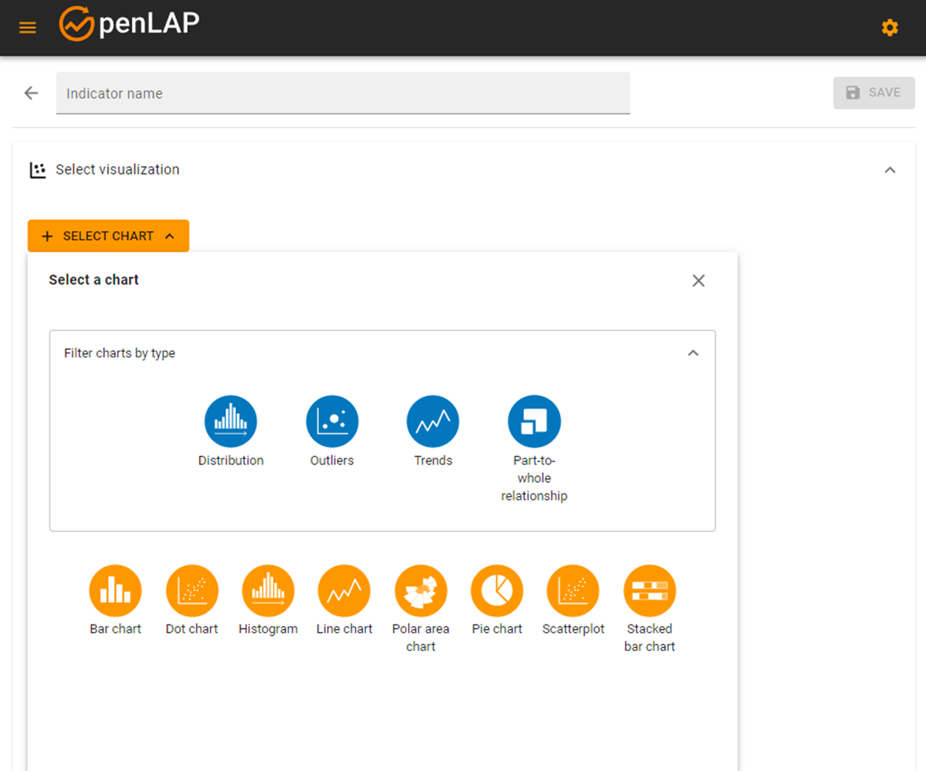}
            \caption{View list of \textit{Tasks}}
            \label{subfig:a4}
        \end{subfigure}
        \\
        \begin{subfigure}[normla]{0.48\textwidth}
            \includegraphics[width=1\textwidth]{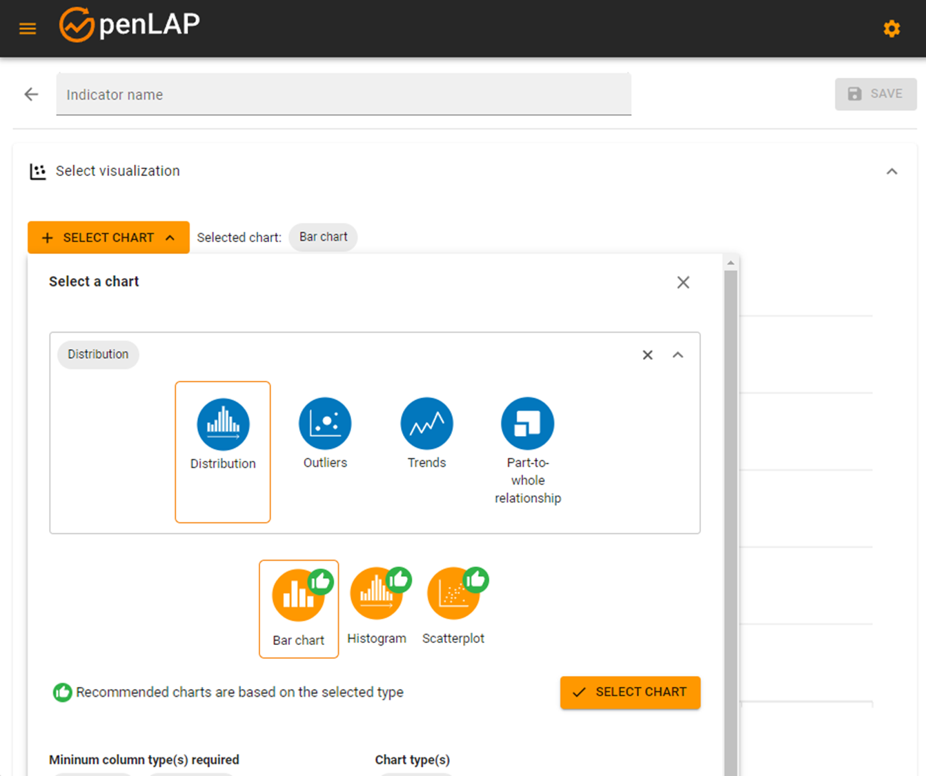}
            \caption{Get recommendations of \textit{Idioms} based on \textit{Task}}
            \label{subfig:a5}
        \end{subfigure}
        ~
        \begin{subfigure}[normla]{0.48\textwidth}
            \includegraphics[width=1\textwidth]{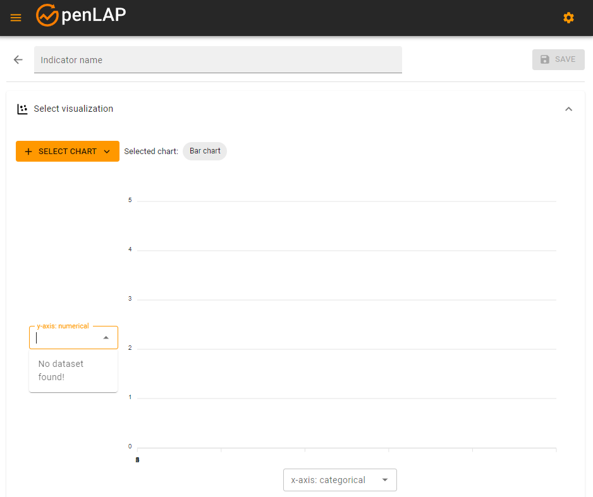}
            \caption{Select \textit{Data}}
            \label{subfig:a6}
        \end{subfigure}
    \end{center}
    \caption{Task-driven approach}
    \label{fig:task-driven-approach}
\end{figure}

The main view of the ISC UI displays the dashboard, as shown in Figure~\ref{subfig:a1} \textbf{(DG1)}. Here, the users can view a list of their custom indicators in form of cards. Users can create a new indicator by clicking on the \textit{Create Indicator} button, which will direct them to a new page, as shown in Figure~\ref{subfig:a2}. Next, the users can either provide a name to their indicator, or select a visualization or data. This provides the users with the flexibility to start their interaction from different parts of the ISC UI \textbf{(DG2)}. If the users click on the accordion panel \textit{Select Visualization}, they will be presented with a list of idioms/charts with illustrations and labels, to help them recognize the idiom/chart, rather than recalling them from memory \textbf{(DG1)} (Figure~\ref{subfig:a3}). Furthermore, the users are presented with a dropdown filter option, which shows a list of tasks with illustrations and labels as well \textbf{(DG1)} (Figure~\ref{subfig:a4}). Users are able to select a specific task of their choice, and the ISC will recommend a list of idioms/charts that are suitable for the selected task, marked with thumbs-up icon \textbf{(DG1, DG3)}. Next, the users can select an idiom/chart of their choice (Figure~\ref{subfig:a5}), preview the visualization, and then select the data, as shown in Figure~\ref{subfig:a6}.

\subsection{Data-driven Approach}
\begin{figure}[htpb]
    \begin{center}
        \begin{subfigure}[normla]{0.48\textwidth}
            \includegraphics[width=1\textwidth]{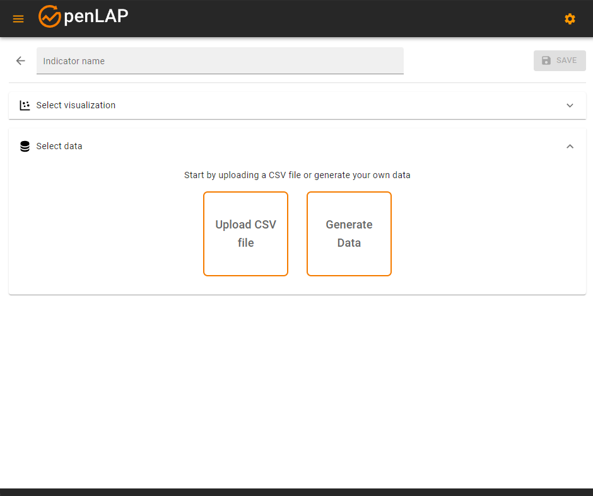}
            \caption{Upload or generate data}
            \label{subfig:b1}
        \end{subfigure}
        ~
        \begin{subfigure}[normla]{0.48\textwidth}
            \includegraphics[width=1\textwidth]{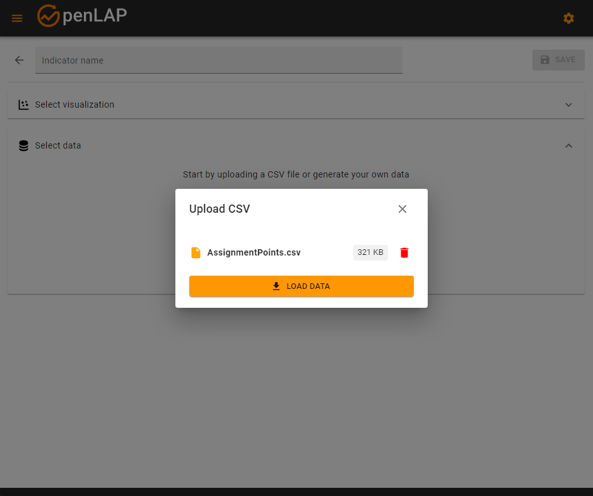}
            \caption{Upload CSV file}
            \label{subfig:b2}
        \end{subfigure}
        \\
        \begin{subfigure}[normla]{0.48\textwidth}
            \includegraphics[width=1\textwidth]{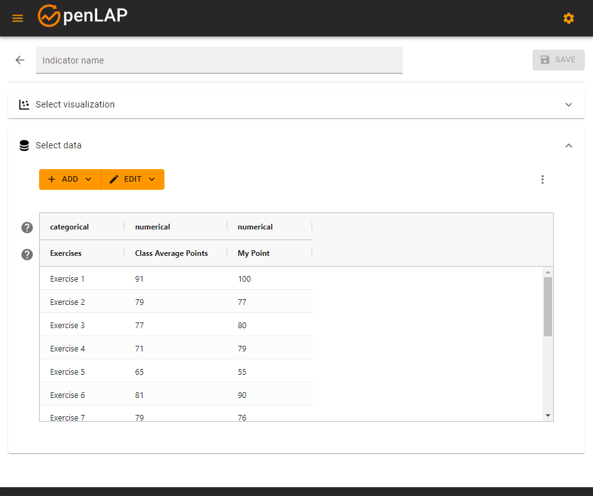}
            \caption{Preview \textit{Data}}
            \label{subfig:b3}
        \end{subfigure}
        ~
        \begin{subfigure}[normla]{0.48\textwidth}
            \includegraphics[width=1\textwidth]{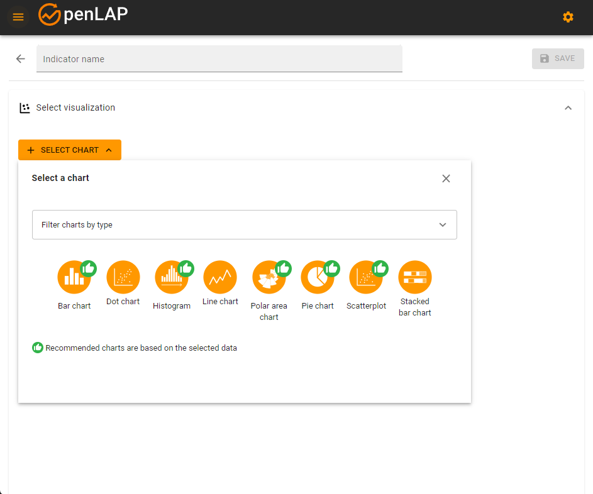}
            \caption{Get recommendations of \textit{Idioms} based on data types}
            \label{subfig:b4}
        \end{subfigure}
        \\
        \begin{subfigure}[normla]{0.48\textwidth}
            \includegraphics[width=1\textwidth]{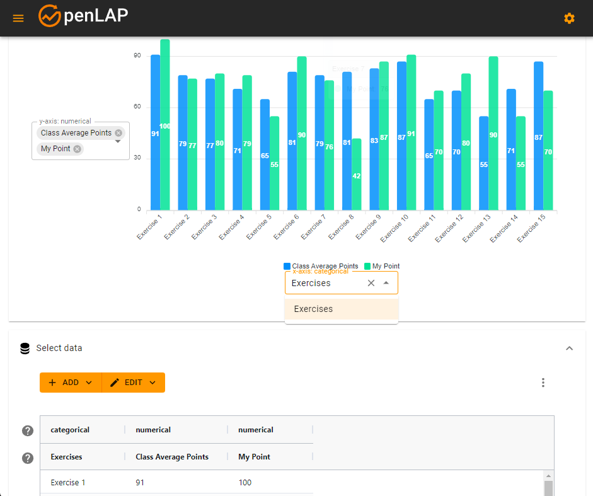}
            \caption{Select appropriate data columns and preview chart}
            \label{subfig:b5}
        \end{subfigure}
    \end{center}
    \caption{Data-driven approach}
    \label{fig:data-driven-approach}
\end{figure}

The users can create a new indicator from the dashboard by clicking on the \textit{Create Indicator} (Figure~\ref{subfig:a1}). Next, the users can click on the accordion panel \textit{Select Data} (Figure~\ref{subfig:a2}), which will present two sets of options, \textit{Upload CSV file} and \textit{Generate Data} \textbf{(DG1)}, as shown in Figure~\ref{subfig:b1}. \textit{Generate Data} will allow users to create a tabular data with column names, set appropriate column types (categorical, categorical (ordered), numerical) and insert rows and its values. Users can also upload a CSV file (Figure~\ref{subfig:b2}), and preview the data in tabular form. As shown in Figure~\ref{subfig:b3}, the top header of the table shows the column types and the second header shows the name of the columns. The ISC is able to detect the type of the column of the uploaded data automatically \textbf{(DG1)}. Next, the users can click on the accordion panel \textit{Select Visualization} to select an appropriate idiom/chart. As shown in Figure~\ref{subfig:b4}, a list of recommended idioms/charts are shown, marked with thumbs-up icon, based on the data types available in the data table \textbf{(DG1, DG3)}. After the users have selected an appropriate idiom/chart of their choice, e.g., Bar chart, they can preview the visualization (Figure~\ref{subfig:b5}). In each of the axes of the Bar chart, the ISC recommends appropriate column types \textbf{(DG3)}. As shown in Figure~\ref{subfig:b5}, the y-axis of the Bar chart requires numerical column types (``Class Average Points'' and ``My Points''), and the x-axis of the Bar chart requires one categorical column type (``Exercises'').

\section{Conclusion and Future Work}
In this paper, we aimed at providing a `quick and dirty' method to allow low cost design of learning analytics (LA) indicators, based on Indicator Specification Cards (ISC). We presented the development details of an intuitive ISC user interface (UI) to support quick and reliable design of LA indicators. 
Further, we proposed two approaches to allow users to be flexible while designing LA indicators, namely a \textit{task-driven approach} and a \textit{data-driven approach}. Future work will focus on the evaluation of the ISC UI with users to investigate the impact on users' perception of (1) control \& personalization, (2) transparency \& trust, and (3) satisfaction \& acceptance of an LA indicator co-design task, facilitated by the ISC UI.

\bibliography{sample-ceur}

\appendix

\end{document}